\documentclass[12pt,preprint]{aastex}

\shorttitle{VVV Survey NIR Photometry of Bulge RR Lyrae Stars}
\shortauthors{D\'ek\'any et al.}

\begin{document}

\title{VVV Survey Near-Infrared Photometry of Known Bulge RR~Lyrae stars: \\the Distance to the Galactic Center and Absence of a Barred Distribution of the Metal-Poor Population}

\author{I.~D\'ek\'any\altaffilmark{1,4}, D.~Minniti\altaffilmark{1,4,5}, M.~Catelan\altaffilmark{1,4}, M.~Zoccali\altaffilmark{1,4}, R.~K.~Saito\altaffilmark{2}, M.~Hempel\altaffilmark{1,4}, O.~A.~Gonzalez\altaffilmark{3}}
\affil{$^1$Instituto de Astrof\'isica, Facultad de F\'isica, Pontificia Universidad Cat\'olica de Chile, Av. Vicu\~na Mackenna 4860, 782-0436 Macul, Santiago, Chile}
\affil{$^2$Departamento de F\'isica, Universidade Federal de Sergipe, Av. Marechal Rondon s/n, 49100-000, S\~ao Crist\'ov\~ao, SE, Brazil}
\affil{$^3$European Southern Observatory, Alonso de C\'ordova 3170, Vitacura, Santiago, Chile }

\altaffiltext{4}{Also at The Milky Way Millennium Nucleus, Av. Vicu\~na Mackenna 4860, 782-0436 Macul, Santiago, Chile}
\altaffiltext{5}{Also at Vatican Observatory, V-00120 Vatican City State, Italy}
\altaffiltext{}{Accepted for publication in Astrophysical Journal Letters}
\altaffiltext{}{Copyright: American Astronomical Society 2013}

\begin{abstract}

We have combined optical and near-infrared data of known RR~Lyrae (RRL) stars in the bulge in order to study the spatial distribution of its metal-poor component by measuring precise reddening values and distances of 7663 fundamental-mode RRL stars with high-quality photometry. We obtain a distance to the Galactic center of $R_0=8.33\pm0.05\pm0.14$~kpc. We find that the spatial distribution of the RRL stars differs from the structures traced by the predominantly metal-rich red clump (RC) stars. Unlike the RC stars, the RRL stars do not trace a strong bar, but have a more spheroidal, centrally concentrated distribution, showing only a slight elongation in its very center. We find a hint of bimodality in the density distribution at high southern latitudes ($b<-5^{\circ}$),
which needs to be confirmed by extending the areal coverage of the current census.
The different spatial distributions of the metal-rich and metal-poor stellar populations suggest that the Milky Way has a composite bulge.

\end{abstract}


\section{Introduction}
\label{introduction}

\noindent RR~Lyrae stars are easily identifiable primary distance indicators, which can be used to trace the structure of the Milky Way. They sample an old and metal-poor population, and are particularly useful in the reddened regions of the Galactic bulge because of several reasons: (i) they have a relatively high number density in the bulge \citep[e.g.,][]{2011AcA....61....1S}; (ii) they follow precise period-luminosity (PL) relations in the near-infrared \citep[NIR; e.g.,][]{2004ApJS..154..633C}; (iii) their PL relations allow the determination of the interstellar extinction on a star-by-star basis; (iv) their metallicities can be estimated using photometric data alone (e.g., \citealt{1996A&A...312..111J,2005AcA....55...59S}); and (v) their brightness and amplitude ($0.1\lesssim a_{K_s}\lesssim0.5$\,mag) make them detectable out to large distances.

The VISTA Variables in the V\'ia L\'actea (VVV) ESO Public Survey \citep{2010NewA...15..433M,2011rrls.conf..145C,2012A&A...537A.107S} aims to probe the 3-dimensional (3D) structure of the Galactic bulge using various distance indicators.
In a series of papers, the VVV $ZYJHK_s$ color atlas was used to trace the spatial distribution of red clump (RC) giants \citep{2011AJ....142...76S,2011ApJ...733L..43M,2011A&A...534L..14G}, and to provide photometric reddening and metallicity maps\footnote{\texttt{http://mill.astro.puc.cl/BEAM/calculator.php}} of the bulge \citep{2011A&A...534A...3G,2012A&A...543A..13G,2013A&A...552A.110G,2013A&A...550A..42C}. The RC stars clearly trace a complex, X-shaped distribution, as was first suggested by \cite{2010ApJ...724.1491M} and \cite{2010ApJ...721L..28N}. They are predominantly metal-rich, with the mode of their [Fe/H] distribution falling between $-0.5$ and $0$~dex, following a radial gradient \citep{2008A&A...486..177Z,2011AAS...21715312J,2013ApJ...765..157J,2012A&A...543A..13G}. However, there is a more metal-poor population of stars present in the bulge. More than 16,000 RR~Lyrae stars have been found in the bulge so far by the OGLE-III survey \citep{2011AcA....61....1S}, and the census is still limited to a fraction of the complete bulge area. Their metallicity distribution is centered around $[{\rm Fe/H}]=-1$ \citep{2012ApJ...750..169P} and has a very narrow spread, suggesting that they trace a more ancient stellar population with respect to the RC stars.

In this Letter, we analyze the spatial distribution of RR~Lyrae stars from the OGLE-III catalog \citep{2011AcA....61....1S} by combining their $I$-band optical data with the available VVV $K_s$-band photometry. Our optical-NIR data set has some important advantages over optical studies.
First, the PL relations have increasing precision and decreasing metallicity dependence towards longer NIR wavelengths \citep{2004ApJS..154..633C}.
The combined optical--NIR data allow us to rely primarily on the highly precise $K_s$-band relation --- with the $I$-band one appearing only in the reddening estimation --- and to leave out the very rough correlation between the absolute $V$-band magnitude and the metallicity from the analysis. 
This yields more precise distances by increasing the precision of both the absolute magnitudes and individual reddening values through the intrinsic color indices.
Secondly, since NIR wavebands are much less sensitive to interstellar reddening than optical ones, and due to the wide wavelength separation between the $I$ and $K_s$ bands, the uncertainties in the intrinsic colors of the stars will cause much smaller errors in the reddening values ($R_{K_s,I-K_s}<<R_{I,V-I}$), and in turn, the distances. Finally, the VVV $K_s$-band average magnitudes are more accurate than the OGLE $V$-band ones because of the higher phase coverage and smaller amplitude of the light-curves --- increasing the accuracy of the distance determination.

\section{The VVV NIR Photometry of the OGLE RR~Lyrae stars}
\label{vvvphot}

Our study is based on VVV $K_s$ photometric data provided by the VISTA Data Flow System \citep[VDFS, see][]{2004SPIE.5493..401E}, acquired between October 19, 2009, and September 28, 2012. We used magnitudes measured on single {\em detector frame stacks} (a.k.a. ``{\em pawprints}'') by aperture photometry, using small, flux-corrected apertures, suitable for 
moderately crowded stellar fields \citep{2004SPIE.5493..411I}. The number of observational epochs varies between $\sim20$ and $60$, depending on the brightness and the location of the object. The limiting $K_s$ magnitude ranges from $\sim 18.0$ to $\sim 16.5$~mag (depending on the Galactic latitude), which is enough to detect RR~Lyrae stars all the way through the Galaxy, even with high interstellar extinction.

We performed a positional cross-matching procedure between sources extracted from VVV images and the complete catalog of 11756 fundamental-mode RR~Lyrae (type RRab) stars from the OGLE-III survey \citep{2011AcA....61....1S} using the \textsc{STILTS}\footnote{\texttt{http://www.star.bristol.ac.uk/$\sim $mbt/stilts/}} software. In order to rule out photometric contamination by seeing-dependent source merging, nearby saturated stars, spurious signals, etc., we excluded all data points with cross-matching separations larger than $0.35''\sim 1$ pixel (less than $10\%$ of all data) or lying closer to the detector edges than the jitter offset, 
and only considered sources with morphological classifications of $-1, -2$ or $1$ \citep[see][]{2012A&A...537A.107S}. The procedure resulted in our {\em initial database} of 9694 $K_s$-band light-curves with OGLE-III $I$-band counterparts.

\begin{figure}
\epsscale{.80}
\plotone{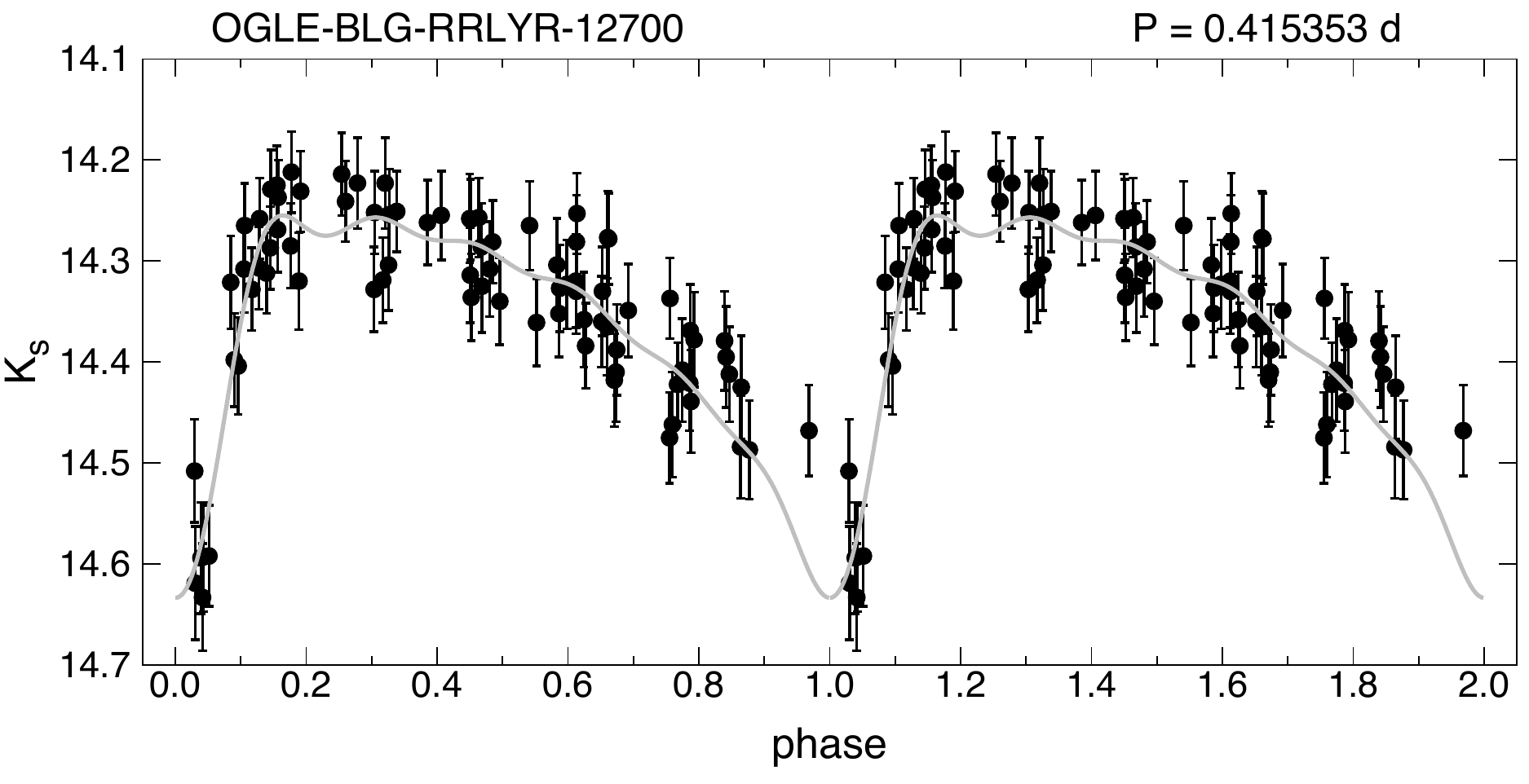}
\caption{VVV $K_s$-band light-curve of a bulge RRab star with the fitted template.
The identifier and period of the star are shown in the header.}
\label{rrab}
\end{figure}

We computed precise mean $K_s$-band magnitudes for each star by employing the template Fourier fitting method of  \cite{2007A&A...462.1007K}, using linear scaling of 6th-order Fourier-sums fitted to the template light-curves of \cite{1996PASP..108..877J}, and iteratively omitting outliers during the procedure. Fig.~\ref{rrab} illustrates our data quality showing a typical light-curve with the fitted template.

\begin{figure}
\epsscale{.80}
\plotone{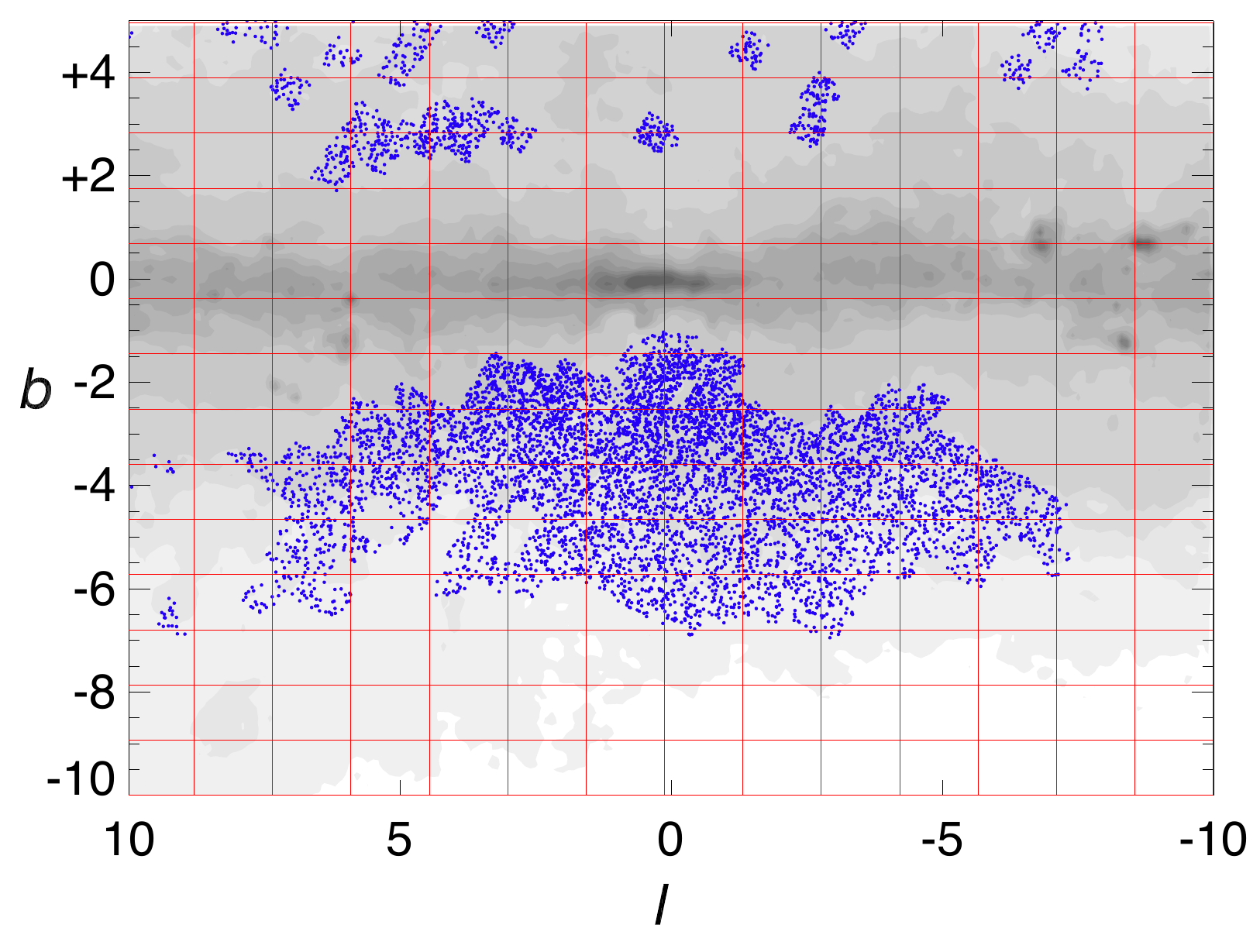}
\caption{
Galactic coordinates of our {\em final sample} of 7663 OGLE-III RRab stars {\em(dots)} in the bulge area of the VVV Survey. The VVV bulge subfields are shown by an overlaying grid. The grayscale background shows the interstellar extinction map of \cite{1998ApJ...500..525S}.}
\label{o3map}
\end{figure}

In order to reject light-curves dominated by noise or systematic variations, we rejected all light-curves with $a\sqrt{N} / \sigma<25$, where $a$ is the total amplitude of the template fit, $\sigma$ is the standard deviation of the residual, and $N$ is the number of data points. The above limit was set by a visual inspection of $\sim10\%$ of all fitted light-curves, and resulted in the rejection of $\sim20\%$ of light-curves in our {\em initial sample}. For minimizing the presence of systematic errors in the computed distances, we required the light-curves to comply with the following set of further selection criteria. We rejected all light-curves having a phase coverage\footnote{1 minus the maximum phase lag} less than $0.6$, and $\sigma>0.1$\,mag. Furthermore, to exclude sources suspected to be affected by blending or fitting runaways, we omitted stars outside the amplitude threshold of $0.08<a_{K_s}<0.6$. By applying these cuts, we obtained our {\em final sample} of 7663 RRab light-curves. Figure~\ref{o3map} shows the location of the these sources in comparison with the bulge area covered by the VVV Survey. The various cuts in the sample does not result in any bias in the ($l,b$) distribution of the objects, and yields an increased relative precision in the traced spatial distribution out to high distances. The cost of this gain is a slightly increased incompleteness towards the far side of the bulge, due to the preference for rejecting faint stars, since their light-curves are more noisy and have less data.

\section{Distances to the RR~Lyrae stars}
\label{distances}

We derived the absolute $K_s$ and $I$ magnitudes of the stars using the theoretical period-luminosity-metallicity (PLZ) relations from \cite{2004ApJS..154..633C}, estimating the apparent equilibrium brightnesses of the stars with their intensity-averaged mean magnitudes obtained from the template-fitting procedure (see Sect.~\ref{vvvphot}). Since the original $K$-band relation in the above study is given in the Glass system, first we transformed the $K$ magnitudes for each and every individual synthetic star in the \cite{2004ApJS..154..633C} synthetic horizontal-branch populations into the VISTA filter system. For this purpose, we used Eqs.~(A1)-(A3) from \cite{2001AJ....121.2851C} to transform first into the 2MASS system, and then the equations provided by CASU\footnote{\tt http://casu.ast.cam.ac.uk/surveys-projects/vista/technical/\newline photometric-properties/sky-brightness-variation/view} for transforming these into the VISTA system. We then recomputed all the linear regressions, leading to the following relation for the VISTA $K_s$ band:
\begin{equation}
M_{K_s} = -0.6365 - 2.347 \log{P} + 0.1747 \log{Z}\,.
\end{equation}
In estimating the $\log Z$ values of the RRab stars, we used the exact same approach as \cite{2012ApJ...750..169P}, applying the empirical $P,\phi_{31}\rightarrow{\rm[Fe/H]}$ relation of \cite{2005AcA....55...59S} with the periods and phase differences derived from the OGLE-III $I$-band light-curves \citep[as provided by][]{2011AcA....61....1S}.

We determined the color excess of each star by comparing its observed color index to the intrinsic one predicted by the theoretical PLZ relations:
\begin{equation}
{\rm E}(I-K_s)=(I-K_s)-(M_I-M_{K_s})~.
\end{equation}
For converting the color excess values into absolute extinction in the $K_s$ band, we adopted the reddening law of \cite{1989ApJ...345..245C} with $R_V=3.1$. Using their Eqs. (2a-b) and (3a-b) with an effective wavelength of $\lambda_{\rm eff}=0.798\,\mu {\rm m}$ for the Cousins $I$-band and $\lambda_{\rm eff}=2.149\,\mu {\rm m}$ for the VISTA $K_s$-band, we obtained the following relation:
\begin{equation}
A_{K_s}=0.164\,E(I-K_s)\,.
\end{equation}
We note that due to the large wavelength difference between the bands, the coefficient in this relation changes only marginally (to the value of $0.160$) if we adopt $R_V=2.5$, as recently suggested by \cite{2013ApJ...769...88N}.

Finally, we calculated the individual distances to our bulge RRab sample using the relation
\begin{equation}
\log{R}=1+0.2\,(K_{s,0}-M_{K_s})\,,
\end{equation}
where $K_{s,0}$ is the extinction-corrected $K_s$-band magnitude.

The precision of the resulting distances to individual stars is significantly higher than in earlier studies due to the advantages of NIR photometry mentioned in Sect.~\ref{introduction}. The typical accuracy of the average $K_s$ magnitudes obtained from the template fitting procedure is between $0.01-0.02$~mag, depending on the amplitude and the brightness of the star --- which is similar to the 0.02~mag nominal accuracy of the $I$-band average magnitudes. The internal dispersion of the photometric metallicity estimator is 0.18\,dex \citep{2005AcA....55...59S}, and its observational error is dominated by the error in $\phi_{31}$, which is determined by the quality of the $I$-band light-curves, and which is negligible for most of the stars in our sample, according to \cite{2012ApJ...750..169P}. The error in $\log{Z}$ propagates into the distance through both $M_{K_s}$ and $A(K_s)$, but the latter gives only a marginal contribution due to the very low ratio of absolute to selective extinction. Assuming a median statistical error of 0.18\,dex in $\log{Z}$, and 0.02~mag for the error in the average magnitudes in both bands, we get a statistical error of only 0.01~mag in $A(K_s)$. The errors propagate into a statistical uncertainty of only 0.053~mag in the extinction-corrected distance moduli, which converts into a distance error of 0.2~kpc for a star at a distance of 8.5~kpc. This is less than half the median statistical error in the distances obtained from optical photometry, as quoted by \cite{2012ApJ...750..169P}. 

In addition to the statistical errors discussed above, there might be a significant net systematic error in the distances, originating from various different sources. A detailed exploration of the possible sources of systematic errors is out of the scope of this Letter; here we merely use a relevant globular cluster study as reference.
\cite{2011MNRAS.416.1056C} provide a detailed comparison of various recent PL/PLZ relations through the distance determination of the globular cluster M5, based on a sizable sample of RR~Lyrae stars; and the resulting distance moduli from different empirical and theoretical relations are within $1\sigma$ agreement, showing a high reliability of the method. Since this cluster is located at a similar distance as the Galactic Center, the spread of 0.07~mag in the distance moduli obtained from 4 different relations for RRab stars in their study is highly indicative of the possible mean systematic error in our analysis. Therefore, we adopted a mean systematic distance error of 0.14~kpc, corresponding to the same error in the distance modulus at the distance of the GC (see Sect~\ref{gcdist}).

\section{The Distance to the Galactic Center}
\label{gcdist}

\begin{figure}
\epsscale{.80}
\plotone{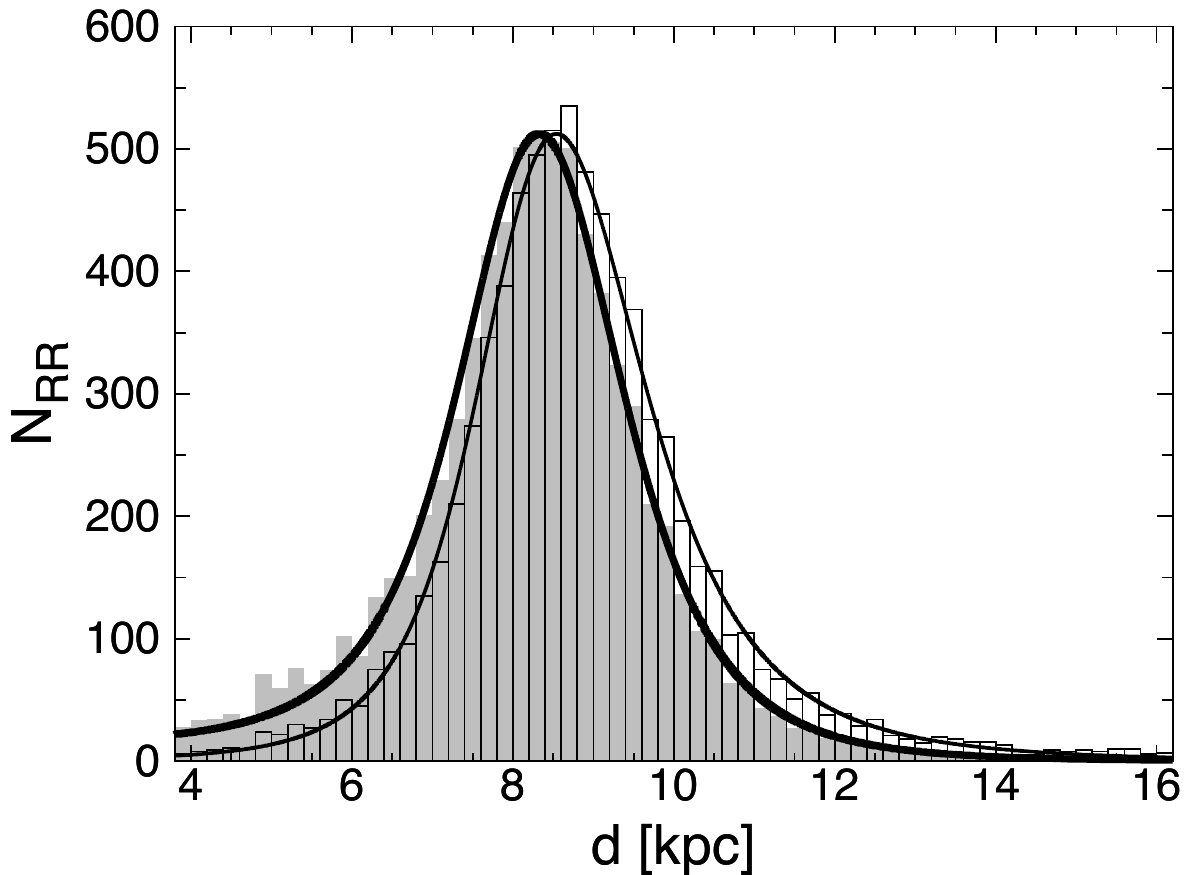}
\caption{Histogram of the projected distances of RRab stars in our {\em final sample}. Empty bars show the histogram of the measured distances, gray bars denote the histogram corrected for the ``cone-effect'' (see text). The thin and thick lines show the density function of the skewed Student's $t$-distribution fitted to the observed distances, before and after applying the correction for the ``cone-effect'', respectively.}
\label{galcen}
\end{figure}

We determined the center of the RR~Lyrae distribution by first projecting the individual distances to the Galactic plane, and then correcting the distance distribution for the distorting effect of our sample being distributed in a solid angle, thus observing more objects at larger distances (i.e., we scaled the distribution by $R^{-2}$). Figure~\ref{galcen} shows the histogram of the projected distances of all RRab stars in our {\em final sample} based on the combined $IK_s$ photometry, before and after correcting for the aforementioned ``cone-effect''. We measured the center of the RR~Lyrae spatial distribution (assumed to coincide with the Galactic Center, $R_0$) by determining the center of the density function of the best-fitting analytical distribution. We found that the distance distribution shows a very significant non-normality due to its heavy tails, and can be best represented by the Student's $t$-distribution. This is consistent with the result of \cite{2012ApJ...750..169P}, who found that the Cauchy-Lorentz\footnote{Student's $t$-distribution with 1 degree of freedom} distribution fits best the distances obtained from the optical data.

The distribution of RR~Lyrae distances shows a natural asymmetry towards higher distances due to the ``cone-effect''. In our case, the distribution is further perturbed by our rejection procedure (see Sect.~\ref{vvvphot}), resulting in a slight incompleteness at higher distances (since the light-curves of fainter stars have generally lower quality). The non-symmetry of the observed distance distribution has to be taken into account in the distribution fitting, in order not to bias the measurement of $R_0$. We fitted a skewed Student's $t$-distribution to the projected distances of our {\em final sample} by likelihood inference \citep[see][]{2003JRSS...65..367}.
Using the fitted parameters, we evaluated the corresponding density function on a dense grid of quantiles, and corrected for the ``cone-effect'' by scaling the obtained synthetic density distribution with $R^{-2}$. Then we determined $R_0$ by numerically computing the location of the maximum of the corrected synthetic distribution. The fitted density curves before and after scaling are shown in Fig.~\ref{galcen}. We obtained a value of $R_0=8.33\pm0.05\pm0.14$\,kpc for the distance to the Galactic Center, where the first error is purely statistical, representing the standard error in the location parameter of the fitted distribution, and the second one is  the estimated systematic error discussed in Sect.~\ref{distances}.
Our measurement of $R_0$ is consistent with recent literature values obtained from various other methods \cite[e.g.,][]{2010RvMP...82.3121G}, and we note the remarkable agreement with the recent result of $8.33\pm0.35$\,kpc\footnote{includes systematic errors}, obtained by \cite{2009ApJ...692.1075G} from monitoring stellar orbits around the central black hole.

\section{The spatial distribution of RR~Lyrae stars}

\begin{figure}
\epsscale{.50}
\plotone{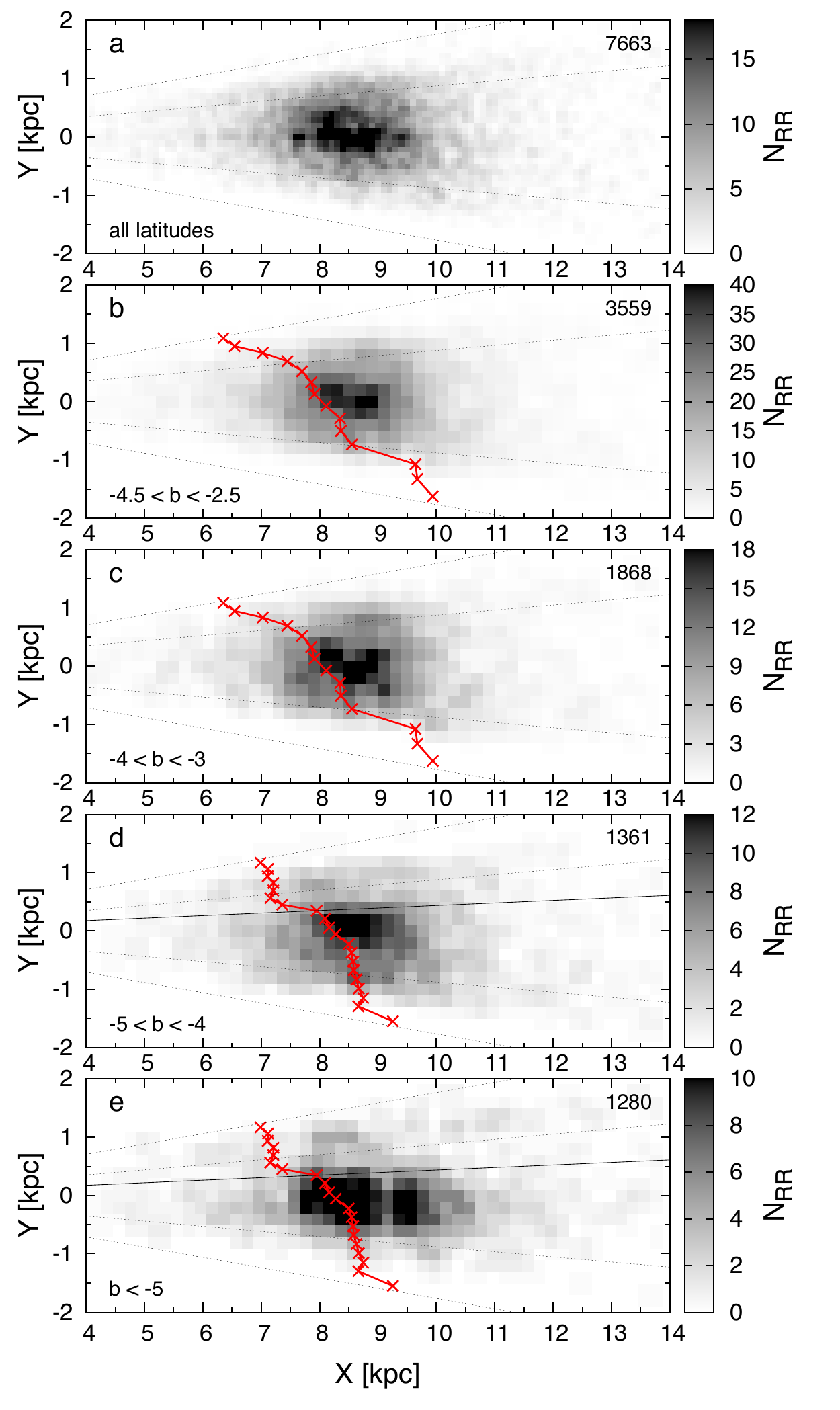}
\caption{Number density diagrams showing the spatial distribution of the RRab stars in our {\em final sample} in various latitudinal stripes, projected onto the Galactic plane. The corresponding latitudinal ranges (in degrees) and the number of stars within them are shown in the bottom left and top right corners of each panel, respectively. Dotted lines denote lines of sight corresponding to $l=\pm5^\circ$ and $\pm10^\circ$, the solid lines in panels {\bf d} and {\bf e} mark the upper limit of contiguous areal coverage in the corresponding stripes (see text). Crosses connected with broken lines in panels {\bf b-e} show the observed location of the long bar at $b=-1^\circ$ ({\bf b, c}) and $b=-5^\circ$ ({\bf d, e}) as found by \cite{2011A&A...534L..14G,2012A&A...543A..13G} based on RC distances. The bin size is 100\,kpc in panel {\bf a} and 200 kpc in panels {\bf b-e}.}
\label{map_hess}
\end{figure}

The distribution of RC giants in the bulge is known to be barred \citep{1997ApJ...477..163S,2013arXiv1303.6430C}. Figure~\ref{map_hess} shows the spatial distribution of the RRab stars obtained in the present study at various latitudinal stripes at negative latitudes, where our sample has sufficiently high number density. The stars show a centrally concentrated distribution at all latitudes. Figure~\ref{map_hess}a shows all stars in our {\em final sample} -- note that the elongation along the line of sight (LOS) is an observational bias due to our incomplete longitudinal coverage. It already implies, however, that the RR~Lyrae stars do not trace an inclined prominent bar like the RC stars do. Figures~\ref{map_hess}b-e show latitudinal subsets of the distribution. The longitudinal coverage between $l=\pm5^\circ$ is very close to contiguous at $-4.5^\circ<b<-2.5^\circ$ and fully contiguous at $-4^\circ<b<-3^\circ$ (cf.~Fig.~\ref{o3map}), therefore the distribution is free of any significant bias between these longitudes. The longitudinal ranges of contiguous coverage are more confined at higher southern latitudes, but should still be free from a significant bias between $l=+2.5^\circ$ and $-5^\circ$ approximately (Figs.~\ref{map_hess}d,e).

In Figs.~\ref{map_hess}b-e, we compare the spatial distributions with the observed orientation of the Galactic bar traced by RC stars at different southern latitudes \citep[data are taken from][see their Figs.~3 and 10, respectively]{2011A&A...534L..14G,2012A&A...543A..13G}. It is immediately visible that the spatial distribution of RR~Lyrae stars is not significantly barred at any of these latitudes. At latitudes $-4.5^\circ<b<-2.5^\circ$, the distribution shows a slight elongation within the inner 1~kpc. This small substructure is inclined by only $-12.5^{\circ}\pm0.5^{\circ}$ with respect to the LOS towards the GC, obtained by fitting elliptical centroids to the distribution with 0.2~kpc binning over a grid of small (X,Y) offsets of the bin positions. Our findings are in contrast with the results of \cite{2012ApJ...750..169P}, that showed a more inclined and longer bar-like substructure at certain negative latitudes based on the study of $V,I$ optical data from the OGLE-III survey. We note again however, that our combined optical-NIR data has superior statistical quality due to the various arguments discussed in Sect.~\ref{distances}, providing extinction and distance values with significantly higher relative precision.

The spatial distribution at higher southern latitudes ($b<-5^\circ$) has a significantly wider peak between distances of 7.7 and 9.7 kpc, approximately. It also shows a hint of bimodality, having a secondary maximum in the number density distribution at a distance of about $9.6$\,kpc. 
It is somewhat resemblant of the split distribution we see in the case of RC stars due to the split arms of the X-shaped structure towards more negative latitudes.
However, the dip test of \cite{1985AStat...13..70}, indicates only marginally significant bimodality ($2\sigma$).
Unfortunately, the current census does not extend to sufficiently high southern latitudes, preventing us from getting a true picture on this possible substructure; a larger sample of RR~Lyrae from a more extended bulge area is required.

\section{Conclusions}

We combined NIR (VVV) and optical (OGLE-III) photometric data in order to measure precise reddenings and distances for bulge RR~Lyrae stars with high-quality light-curves. Our data are only little affected by the large and non-uniform interstellar extinction in these regions, allowing us to reach higher precision with respect to optical studies. We derive a distance to the Galactic Center of $R_0=8.33\pm0.05\pm0.14$~kpc. 
We find that the distribution of fundamental-mode RR~Lyrae stars is different from the known distribution of RC giants. The most striking fact is that the RR~Lyrae stars do not show a strong bar in the direction of the Galactic bulge as do the clump giants, but their distribution is more spherical, with only the central $\sim1$~kpc showing a bar-like substructure.

In recent years, the structure of the bulge has been discovered to be more complex than just a simple bar.
It is now well-established that the predominantly metal-rich bulge RC stars trace an X-shaped structure \citep{2010ApJ...724.1491M,2010ApJ...721L..28N,2011AJ....142...76S,2012A&A...543A..13G}.
Recently, \cite{2012ApJ...756...22N,2013MNRAS.432.2092N} found that this structure disappears from the RC distribution at low metallicities. 
The observed prominent difference between the 3-D structures traced by the RC and metal-poor RR~Lyrae stars is in line with their findings, and suggests that the Milky Way has a composite bulge, retaining an older, more spheroidal bulge component \citep[see, e.g.][]{2003A&A...399..961S,2013ApJ...763...26O}. The existence of a small, central bar-like structure in such a component like the one traced by the RR~Lyrae stars is consistent with current numerical simulations of composite bulges, and is a result of the angular momentum transfer between the bar and the initial classical bulge during their co-evolution \citep{2013MNRAS.430.2039S}.

\acknowledgments

We gratefully acknowledge the use of data from the ESO Public Survey program ID 179.B-2002 taken with the \facility{VISTA} telescope, data products from the Cambridge Astronomical Survey Unit, funding from the BASAL CATA Center for Astrophysics and Associated Technologies through grant PFB-06, and the Ministry for the Economy, Development, and Tourism's Programa Iniciativa Cient\'ifica Milenio through grant P07-021-F, awarded to The Milky Way Millennium Nucleus. We acknowledge the support of FONDECYT Regular grants No. 1130196 (to D.M.), 1110326 (to I.D. and M.C.), and 1110393 (to M.Z.).

\clearpage

\end{document}